\documentclass[fleqn,usenatbib]{mnras}

\usepackage{newtxtext,newtxmath}

\usepackage[T1]{fontenc}

\DeclareRobustCommand{\VAN}[3]{#2}
\let\VANthebibliography\thebibliography
\def\thebibliography{\DeclareRobustCommand{\VAN}[3]{##3}\VANthebibliography}

\usepackage{graphicx}	
\usepackage{amsmath}	

\title[Full shape analysis of (e)BOSS]{Combined full shape analysis of BOSS galaxies and eBOSS quasars using an iterative emulator}

\author[R. Neveux et al.]{\parbox{\textwidth}{
Richard~Neveux\thanks{E-mail: richard.neveux@ed.ac.uk}$^{1}$,
Etienne Burtin$^{1}$,
Vanina Ruhlmann-Kleider$^{1}$,
Arnaud de Mattia$^{1}$,
Agne Semenaite$^{2}$,
Kyle~S.~Dawson$^{3}$,
Axel de la Macorra$^{4}$,
Will J. Percival$^{5,6,7}$,
Graziano Rossi$^{8}$,
Donald P. Schneider$^{9,10}$,
Gong-Bo Zhao$^{11}$
} \vspace*{4pt} \\
\scriptsize $^{1}$ IRFU,CEA, Universit\'e Paris-Saclay, F-91191 Gif-sur-Yvette, France\vspace*{-2pt} \\ 
\scriptsize $^{3}$ Max-Planck-Institut f\"ur Extraterrestrische Physik, Postfach 1312, Giessenbachstr., 85748 Garching bei M\"unchen, Germany\vspace*{-2pt} \\ 
\scriptsize $^{3}$ Department Physics and Astronomy, University of Utah, 115 S 1400 E, Salt Lake City, UT 84112, USA\vspace*{-2pt} \\ 
\scriptsize $^{4}$ Instituto de F\'isica, Universidad Nacional Aut\'onoma de M\'exico, Apdo. Postal 20-364, M\'exico\vspace*{-2pt} \\ 
\scriptsize $^{5}$ Waterloo Centre for Astrophysics, University of Waterloo, Waterloo, ON N2L 3G1, Canada\vspace*{-2pt}  \\ 
\scriptsize $^{6}$ Department of Physics and Astronomy, University of Waterloo, Waterloo, ON N2L 3G1, Canada\vspace*{-2pt}  \\
\scriptsize $^{7}$ Perimeter Institute for Theoretical Physics, 31 Caroline St. North, Waterloo, ON N2L 2Y5, Canada\vspace*{-2pt}  \\
\scriptsize $^{8}$ Department of Physics and Astronomy, Sejong University, Seoul 143-747, Korea\vspace*{-2pt} \\ 
\scriptsize $^{9}$ Department of Astronomy \& Astrophysics, Pennsylvania State University, University Park, PA 16802, USA\vspace*{-2pt} \\ 
\scriptsize $^{10}$ Institute for Gravitation and the Cosmos, Pennsylvania State University, University Park, PA 16802, USA\vspace*{-2pt} \\ 
\scriptsize $^{11}$ National Astronomy Observatories, Chinese Academy of Science, Beijing, 100012, P.R. China\vspace*{-2pt} \\ 
}
\date{Accepted XXX. Received YYY; in original form ZZZ}

\pubyear{2020}

\begin{document}
\label{firstpage}
\pagerange{\pageref{firstpage}--\pageref{lastpage}}
\maketitle

\begin{abstract}

Standard full-shape clustering analyses in Fourier space rely on a fixed power spectrum template, defined at the fiducial cosmology used to convert redshifts into distances, and compress the cosmological information into the Alcock-Paczynski parameters and the linear growth rate of structure. In this paper, we propose an analysis method that operates directly in the cosmology parameter space and varies the power spectrum template accordingly at each tested point. Predictions for the power spectrum multipoles from the TNS model are computed at different cosmologies in the framework of $\Lambda \rm{CDM}$. Applied to the final eBOSS QSO and LRG samples together with the low-z DR12 BOSS galaxy sample, our analysis results in a set of constraints on the cosmological parameters $\Omega_{\rm cdm}$, $H_0$, $\sigma_8$, $\Omega_{\rm b}$ and $n_s$. 
To reduce the number of computed models, we construct an iterative process to sample the likelihood surface, where each iteration consists of a Gaussian process regression. This method is validated with mocks from N-body simulations.
From the combined analysis of the (e)BOSS data, we obtain the following constraints: $\sigma_8=0.877\pm 0.049$ and $\Omega_{\rm m}=0.304^{+0.016}_{-0.010}$ without any external prior. The eBOSS quasar sample alone shows a $3.1\sigma$ discrepancy compared to the Planck prediction.

\end{abstract}




\section{Introduction}

Large scale structures (LSS) bring essential information on how the Universe evolved through the last $10~\mathrm{Gyr}$. The current spectroscopic surveys allow the construction of 3D maps of the Universe 
using biased tracers of matter (e.g. galaxies or quasars). In standard clustering analyses, the information 
encoded in the power spectrum of the tracer distribution 
in redshift space is compressed into three parameters: the Alcock-Paczynski parameters ($\alpha_\parallel$ and $\alpha_\perp$) that constrain the cosmological distances through the position of the baryon acoustic oscillations features (BAO) and the amplitude of velocity fluctuations $f\sigma_8$ measured from the clustering power as a function of the angle with respect to the line-of-sight.

This compression is accurate but not exact as the information from the shape of the power spectrum is ignored \citep{brieden}. In 
standard analyses,
a fiducial cosmology is 
chosen to compute both the distances from object redshifts and the model power spectrum.
In this case, the cosmological constraints extracted from the data present a dependency in the fiducial cosmology that requires 
to be treated as a systematic error~\citep{smith}.

Recently, the eBOSS collaboration performed BAO and full shape (BAO+RSD) analyses at different epochs of the Universe evolution, 
from $z=0.6$ to $z=2.2$, in Fourier and configuration spaces~\citep{gilmarin,bautista,demattia,tamone,neveux,hou}. The information from each tracer was compressed into a radial distance, a transverse distance and the growth rate of structures at each effective redshift. The eBOSS cosmological analysis~\citep{eboss} was performed by fitting 
cosmological parameters onto these compressed quantities and showed 
that the inferred bias from the dependency in the fiducial cosmology was low enough regarding eBOSS statistical power but has to be considered for next-generation surveys like DESI \citep{desi} or Euclid \citep{euclid}.
 Besides, different techniques have been used to take into account the full shape of the power spectrum, avoiding the compression step to directly constraint base cosmological parameters and accounting for the full shape of the power spectrum \citep{sanchez,grieb,troster,semenaite}, and more recently, using the Effective Field Theory approach \citep{ivanov,amico,colas,chen,zhang}.

In the present analysis, we use a given cosmology to convert redshifts into 
distances to compute the data power spectrum. On the other hand, we evaluate the model power spectrum for different cosmologies in the $\Lambda$CDM framework and compare the measured power spectrum to each model, letting nuisance parameters free while the informative parameters ($\alpha_\parallel$, $\alpha_\perp$ and $f\sigma_8$) take their values as predicted in each considered cosmology.
In this way, we are able to reconstruct the likelihood surface directly in the cosmological parameter space. 
We use the TNS model \citep{tns} where calculation are done in the RegPT framework with correction computed at two loops \citep{Taruya_2012}. 
Computing the model takes around 300 cpu hour per point, which is too 
expensive to be used in a Monte Carlo Markov Chain algorithm. To circumvent this problem, we construct an iterative emulator based on the algorithm 
presented in \cite{pellejero}. Emulation has been gaining 
a lot of interest, as it allows to speed up the analysis by reducing the 
number of expensive function evaluations required by allowing to 
interpolate between model predictions at a set of points in the 
parameter space. This method has been used in a number of galaxy 
clustering analyses to directly emulate observables such as galaxy power 
spectrum \citep{kwan} and correlation function \citep{zhai} or 
model ingredients such as the redshift space power spectrum of halos 
\citep{kobayashi}. Nevertheless, when building an emulator one must 
consider how theory model uncertainty might propagate to the 
cosmological parameter space which is non-trivial and needs to be 
validated for each new statistic or range of scales employed. The 
algorithm presented by \cite{pellejero} aims at dealing 
this challenge by emulating the likelihood function directly, which 
additionally simplifies the process of building the emulator by reducing 
the dimensions of the emulated quantity. It allows us to compute 
100-1000 fewer models to span the cosmological parameter space and 
interpolate the whole likelihood surface by running a Gaussian process 
regression at each iteration. 

In section~\ref{sec:catalogue}, we present the surveys we analysed as well as the mocks used to test our pipeline, 
section~\ref{sec:method} presents the method, and 
section~\ref{sec:results} lists the results obtained on mocks and on the (e)BOSS data.

\section{Catalogues and mocks}
\label{sec:catalogue}

In this section, we present the BOSS and eBOSS data sets used in this work and 
describe the mocks built from the OuterRim N-body simulations used to test the analysis.

\subsection{Data power spectrum}
\label{sec:data}

We analyse hereafter three samples of the SDSS collaboration \citep{blanton}; the QSO and LRG sample of the extended Baryon Oscillation Spectroscopic Survey (eBOSS; \cite{dawson_2016}) and the low-z galaxy sample of BOSS \citep{dawson_2013}. Those data have been taken using the optical spectrograph of BOSS \citep{smee} in the 2.5m Sloan Foundation Telescope \citep{gunn}. The quasars and galaxies target selections are presented in \cite{myers} and \cite{prakash}, respectively. We summarise, here, the statistic of those samples:

\begin{itemize}
    \item[$\bullet$] the eBOSS DR16 low-z quasar sample ($0.8<z<2.2$) including $343\,708$ objects,
    \item[$\bullet$] the eBOSS DR16 LRG sample ($0.6<z<1.0$) including $377\,458$ objects,
    \item[$\bullet$] the BOSS DR12 low-z LRG sample ($0.2<z<0.5$) including $604\,	002$ objects.
\end{itemize}
Our study is performed in Fourier space, and for the data part, we profit from all 
the results provided by the SDSS collaboration~\citep{beutler,gilmarin,neveux}. This includes power spectrum multipole measurements, window functions and fast mock power spectra used to compute the covariance matrix. The three samples are divided into northern (NGC) and southern (SGC) galactic caps. The 
k-ranges used in this work are the same as 
those used in each of the standard analyses and are summarised in table~\ref{tab:k_par_survey}. Notice that the window functions are normalised following the normalisation of the power spectrum to ensure the consistency of the analysis \citep{demattia_cic}.

\begin{table}
	\centering
	\caption{$k$-range used for the 3 (e)BOSS surveys.}
	\label{tab:k_par_survey}
	\begin{tabular}{|c|c|c|} 
	\hline
    Survey & Multipole & $k$-range $\rm~[h\cdot Mpc^{-1}]$\\
\hline
\hline
QSO DR16 & $[0, 2, 4]$ & $[0.02,0.30]$\\
\hline
LRG DR16 & $[0, 2, 4]$ & $[0.02,0.15]$\\
\hline
low-z galaxies DR12 & $[0, 2]$ & $[0.02,0.15]$\\
 & $[4]$ & $[0.02,0.1]$\\
\hline
\end{tabular}
\end{table}

The eBOSS LRG and quasar samples slightly overlap at redshift $0.8<z<1.0$. However, it consists of a relatively small part of the LRG sample. Following eBOSS analysis, the correlation has been estimated to be less than 0.1 and is neglected. For 
BOSS DR12, we use 
only the low-z part of the sample since the highest redshift part is included in the eBOSS DR16 sample. For the sake of simplifying the analysis, the BOSS galaxies within $0.5<z<0.6$ are ignored. 

We use the fast mock power spectra provided with each sample to compute the covariance matrices for the individual likelihood computations, consistently with the individual analyses. Those power spectra are computed assuming the same redshift to distance relation as the data power spectrum. 

\subsection{QSO mocks}
\label{sec:mocks}

To test the analysis, we make use of mocks constructed from the OuterRim N-body simulation~\citep{outerrim}.
This simulation contains $10\,240^3$ dark matter particles of mass $m=1.85.10^9~\mathrm{M_{\sun} h^{-1}}$ in a box of length $L=3h^{-1}Gpc$  with periodic boundary conditions. It 
was produced with a flat $\Lambda \mathrm{CDM}$ cosmology:

\begin{equation}
\begin{split}
h = 0.71, \,\, \Omega_{\rm cdm}\mathrm{h}^2 = 0.1109, \,\, \Omega_{\rm b}\mathrm{h}^2 = 0.02258, \\
\sigma_{8} = 0.8, \,\, \mathrm{n_{s}} = 0.963. \qquad
\end{split}
\label{eq:outerrim_cosmology}
\end{equation}

We use QSO mocks built from one snapshot at $z=1.433$ of the OuterRim simulation as explained in \cite{smith}. We use a halo occupation distribution (HOD) that is a function of the 
halo mass, composed by a top hat function to populate halos with 
central quasars and a power law for satellites. 
This HOD is used as our baseline in the following and is described in more details in \cite{smith} where it is called "mock2". 
It was chosen as it is very well fit by the TNS model described in Sec~\ref{sec:model}. Therefore, we ensure that a 
difference in the cosmological inference 
of the present analysis would be due to our technique and not to the response of the TNS model to this particular HOD. We average the power spectra 
over 100 realisations for this particular HOD.

The covariance matrix is rescaled considering that these 100 realisations 
are independent. As all 
realisations come from a unique dark matter distribution and so are not strictly independent, 
errors may be underestimated.

From the same snapshot, 
we also 
use alternative mocks built with a different HOD, called "mock4" in \cite{smith}. This HOD 
combines a smooth step function for 
central quasars and a power law for 
satellites that corresponds to a more physical HOD for quasars.
\section{Analysis method}
\label{sec:method}

We present here the general pipeline based on~\cite{pellejero} and explain the different steps in more detail in the following subsections.

\begin{enumerate}
\item We compute the data power spectrum for a given fiducial cosmology.
\item We create a Latin Hypercube sampling (LHS) to span the cosmological parameter space efficiently.
\item We compute all cosmology-dependent terms of the model power spectrum for each point of the LHS and each effective redshift of the data using the 2-loop correction TNS model.
\item 
For each point of the LHS, we fit the full model power spectrum to the data, minimizing only over nuisance parameters (as informative parameters are set to their true value in each model). Therefore, we obtain a likelihood value at each point.
\item We use a Gaussian process to interpolate the likelihood surface in the cosmological parameter space.
\item We select new points in the cosmological parameter space using the acquisition function to compute new power spectrum models. 
\item We fit those new model power spectra to the data, minimizing over nuisance parameters, and again run a Gaussian process.
\item The two last points are iterated up to the convergence of the estimate likelihood distribution.
\end{enumerate}

\subsection{Data power spectrum}

We use the data power spectra provided by the SDSS collaboration 
which have been all 
computed within the fiducial cosmology:

\begin{equation}
\begin{split}
h = 0.676, \,\, \Omega_{m} = 0.31, \,\, \Omega_{b}h^2 = 0.022, \\
\sigma_{8} = 0.8, \,\, n_s=0.97 \qquad
\end{split}
\label{eq:fiducial_cosmology}
\end{equation}

\subsection{Latin Hypercube Sampling}

To span the cosmological parameter space efficiently, we use a 5D Latin Hypercube sampling (LHS). Assuming $n$ is the number of cosmological parameters, 
the LHS technique consists of dividing each dimension of the parameter space into $n$ intervals. Then, for each parameter, the $n$ intervals are filled only once, thereby enforcing minimal distance between points compared to a random Poisson sampling. 

The cosmological parameter space chosen for this work 
encompasses large variations in order to avoid 
biasing the analysis:
\begin{equation}
\begin{split}
\Omega_{\rm cdm}=[0.05,0.9], \sigma_8=[0.65,1.5], \Omega_{\rm b}=[0.015,0.066],\\ n_s=[0.8,1.15], H_0=[61,76].
\end{split}
\label{eq:prior_range}
\end{equation}

Those ranges 
can be seen as a set of wide flat priors.

\subsection{Model power spectrum}
\label{sec:model}

To model the power spectrum, we use the RegPT treatment of the TNS model \citep{Taruya_2012}. 
In this framework, the redshift space tracer power spectrum is given by
\begin{align}
P_{\mathrm{t}}(k,\mu) &= D(k,\mu,\sigma_{v},a_{\rm vir}) \left[P_{\mathrm{t},\delta\delta}(k) + 2f\mu^{2}P_{\mathrm{t},\delta\theta}(k) \right. \nonumber \\
& \left. + f^{2}\mu^{4}P_{\theta\theta}(k) + b_{1}^{3}A(k,\mu,f/b_1) + b_{1}^{4}B(k,\mu,f/b_1)\right],
\label{eq:power_galaxy_rsd}
\end{align}
where the wavenumber $k$ is the norm of the wavevector $\mathbf{k}$ and $\mu$ its cosine angle with respect to the line-of-sight. 
$f$ is the linear growth rate of structure.

Here, $P_{\mathrm{t},\delta\delta}(k)$ and $P_{\mathrm{t},\delta\theta}(k)$ are the tracer-tracer and tracer-velocity power spectra, respectively, 
and are given by:
\begin{align}
P_{\mathrm{t},\delta\delta}(k) & = b_{1}^{2}P_{\delta\delta}(k) + 2b_{2}b_{1} P_{b2,\delta}(k) + 2b_{s2}b_{1}P_{bs2,\delta}(k) \nonumber \\
& + 2b_{3\mathrm{nl}}b_{1}\sigma_{3}^2(k)P_{\mathrm{m}}^{\mathrm{lin}}(k) + b_{2}^{2}P_{b22}(k) \nonumber \\
& + 2b_{2}b_{s2}P_{b2s2}(k) + b_{s2}^{2}P_{bs22}(k) + N_{g},
\label{eq:power_galaxy_galaxy}
\end{align}
and:
\begin{align}
P_{\mathrm{t},\delta\theta}(k) &= b_{1}P_{\delta\theta}(k) + b_{2}P_{b2,\theta}(k) \nonumber \\
& + b_{s2}P_{bs2,\theta}(k) + b_{3\mathrm{nl}}\sigma_{3}^{2}(k)P_{\mathrm{m}}^{\mathrm{lin}}(k),
\label{eq:power_galaxy_velocity}
\end{align}
where $b_1$ is the linear bias, $b_2$ the second order local bias, $b_{s2}$ the second order non-local bias, $b_{3nl}$ the third order non local bias and $N_g$ the constant stochastic term. Assuming local Lagrangian bias, and following \cite{saito}, we set the non-local biases to:
\begin{align}
b_{s2} &= -\frac{4}{7}\left(b_{1} - 1\right), \\
b_{3\mathrm{nl}} &= \frac{32}{315}\left(b_{1} - 1\right).
\end{align}
We also assume no velocity bias, $P_{t,\theta\theta}=P_{\theta\theta}$. 

In this work, at each point in the cosmological parameter space we compute the linear matter power spectrum with \texttt{CLASS} 
and evaluate the non-linear matter power spectra $P_{\delta\delta}$,$P_{\delta\theta}$ and $P_{\theta\theta}$, as well as RSD correction terms $A(k,\mu,f/b_1)$ and $B(k,\mu,f/b_1)$ at the 2-loop order, these last terms taking most of the computing time. The $1$-loop bias terms $P_{b2,\delta}(k)$, $P_{bs2,\delta}(k)$, $\sigma_{3}^2(k)$, $P_{b2,\theta}(k)$ and $P_{bs2,\theta}(k)$ are provided in \cite{beutler}.

Finally, to account for 
non-linear effects, we include an overall damping function,
\begin{equation}
    D(k,\mu,\sigma_{v},a_{\rm vir})=\frac{1}{\sqrt{1+(k\mu a_{\rm vir})^2}}\exp{\left[-\frac{(k\mu \sigma_v)^2}{1+(k\mu a_{\rm vir})^2}\right]}.
	\label{eq:fog}
\end{equation}
As explained in \cite{neveux}, this prescription allows to dissociate the Gaussian and the non-Gaussian part of the small scale non-linear effects, 
through the parameters $\sigma_v$ and $a_{\rm vir}$, respectively. This damping term stands for the Finger-of-God effect as well as the redshift uncertainty. This damping term is based on the Finger-of-God of \cite{sanchez}, where we removed the $f$-dependency. Such an empirical term has been shown to correctly model various redshift smearing schemes in the eBOSS mock quasar challenge~\citep{smith}. Altogether, the model power spectrum is thus described by 5 nuisance parameters ($b_1$, $b_2$, $N_g$, $\sigma_v$ and $a_{\rm vir}$).

We make use of the model implementation of \cite{demattia}, which has been tested against mocks as part of the eBOSS mock challenge \citep{alam,smith} for redshifts $z=0.859$ and $z=1.433$. Those tests state that the main systematic error arises when using a template model cosmology different from the truth. Considering the tests when cosmologies match, the mock challenge indicates errors smaller than 3\% for $f\sigma_8$ and 0.5\% for the $\alpha$ at both redshifts.
Note that the model power spectrum requires to include the survey window function which is computed with the same fiducial cosmology as that used to estimate the data power spectrum.

\subsection{Scaling the power spectrum}

In standard clustering analyses, the same cosmology is used for the fiducial cosmology and the power spectrum model, and the scale factors are left free in the fitting procedure to recover the underlying cosmology. In this work, each template power spectrum model is evaluated in a cosmology that is different from the fiducial cosmology used to convert redshifts into distances. This difference produces geometrical distortions evaluated through the Alcock-Paczynski (AP) test \citep{alcock} that are exactly accounted for by introducing two scaling factors:

\begin{equation}
\label{eq:expected_dilationscale}
\alpha_{\parallel}=\frac{D_{\rm H}^{\rm true}(z)}{D_{\rm M}^{\rm fid}(z)},\qquad 
\alpha_{\perp}=\frac{D_{\rm M}^{\rm true}(z)}{D_{\rm M}^{\rm fid}(z)},
\end{equation}

for the line-of-sight and the transverse components, respectively, with distances evaluated in $\rm Mpc$.
Then, the wavenumber observed with the fiducial cosmology $k$ and the true wavenumber $k'$ are related by $k_{\|}'=k_{\|}/\alpha_{\|}$ and $k_{\perp}'=k_{\perp}/\alpha_{\perp}$. 
Transferring this into $k=\sqrt{k_{\|}^2+k_{\perp}^2}$ and $\mu$ the cosine with the line-of-sight 
leads to:

\begin{equation}
\begin{split}
k&'=\frac{k}{\alpha_{\perp}}\left[1+\mu^2\left(\left(\frac{\alpha_\perp}{\alpha_\|}\right)^2-1\right)\right]^{1/2}\\
\mu'&=\mu{\frac{\alpha_\perp}{\alpha_\|}}\left[1+\mu^2\left(\left(\frac{\alpha_\perp}{\alpha_\|}\right)^2-1\right)\right]^{-1/2}.
\\
\end{split}
\end{equation}


The above $(k, \mu) \rightarrow (k', \mu')$ mapping must be used together with Equation~\ref{eq:power_galaxy_rsd} to include the AP effect in the model power spectrum to be compared to data (see Equation 19 of~\cite{demattia}).


In this work, we operate directly in the $\Lambda \mathrm{CDM}$ framework and the likelihood surface we build is expressed as a function of 5 cosmological parameters,  $\Omega_{\rm cdm}, \sigma_8, n_s, \Omega_{\rm b}$ and $H_0$, where $\sigma_8$ is the normalisation of the linear power spectrum at redshift $z=0$. Values of the standard parameters $\alpha_{\perp,\parallel}$ and $f\sigma_8(z)$ at any redshift of interest 
can be inferred from those of the cosmological parameters, using  equation~\ref{eq:expected_dilationscale} for the dilation scales and taken from CLASS for the linear growth rate of structure $f$. 

\subsection{Fit of nuisance parameters}

For each model under test and for each galactic cap of each tracer, we find the best-fit nuisance parameters by maximizing the likelihood function,

\begin{equation}
L \propto e^{-\chi^2/2},
\end{equation}
with
\begin{equation}
\chi^2=(P_{\rm data}-P_{\rm model}(\theta))^tW(\rm P_{data}-P_{\rm model}(\theta)),
\end{equation}

where $P_{\rm data}$ is the data vector of power spectrum multipoles and $P_{\rm model}$ is the corresponding vector for the model that is a function of the parameters. We perform the $\chi^2$ minimisation using the code MINUIT. The inverse of the covariance matrix, $W$, is computed from mocks and corrected for the finite number of mocks following the \cite{hartlap} prescription:
\begin{equation}
    W=\frac{N-n-2}{N-1}C^{-1},
\end{equation}
with $N$ the number of mocks used in the construction of the covariance matrix and $n$ the size of the data vector. Following \cite{Percival+2014} and in line with standard analyses, we apply an additional scaling on the cosmological parameter covariance that results in an increase of the errors of the order of a few per cent. 

Then, for each point of the parameter space under consideration, 
the final data likelihood is obtained by taking the product of likelihoods for each tracer (QSO, LRG) and galactic caps (NGC, SGC), maximising its value over the five free nuisance parameters of the model (see section~\ref{sec:model}) for each tracer. 

\subsection{Gaussian process}
\label{sec:gp}

To interpolate the likelihood surface in the 5-dimensional cosmological parameter space, we use a non-parametric technique that gives us the expected likelihood value at every point of the parameter space and the error on this interpolation. Two hypotheses must be verified to perform an efficient Gaussian process. The interpolated points have to be within the convex hull formed by the computed points. We use the approximation of restricting the interpolation interval by 5\% for all parameters.
The second hypothesis is that the surface to interpolate must be smooth enough; a small deviation in the hyper-parameter space must induce a small deviation in the function. We expect the log-likelihood surface to follow this statement.

We use a Gaussian process to estimate the value of the logarithm of the likelihood at a point (in the cosmological parameter space) $\boldsymbol{y}$, $f_y ={\rm log}L(\boldsymbol{y})$, knowing the value of the likelihood on a set of points $X~=~\{\boldsymbol{x}_1,\boldsymbol{x}_2,...,\boldsymbol{x}_N\}$, $\boldsymbol{f}_X ={\rm log}L(X)$. We must evaluate the probability distribution 
of $f_y$ knowing $\boldsymbol{f}_X$; $p(f_y|\boldsymbol{f}_X,X,y)$. 
Assuming that the joint distribution of $\boldsymbol{f}_X$ and $f_y$ is Gaussian,
\begin{equation}
    p(\boldsymbol{f}_X,f_y)=\mathcal{N}
\begin{pmatrix}
\begin{pmatrix}
\boldsymbol{\mu}_x \\
\mu_y
\end{pmatrix}
,
\begin{pmatrix}
K_X & K_{Xy} \\
K_{Xy}^T & K_y.
\end{pmatrix}
\end{pmatrix}
\end{equation}
As in a standard Gaussian emulator analysis, we assume the central values of the multivariate Gaussian distribution to be $\mu_x=\mu_y=0$. The interpolation is entirely governed by the covariance terms, $K_X$ referring to the covariance of $\boldsymbol{f}_X$, $K_y$ the (scalar) covariance of $f_y$, and $K_{Xy}$ the (vector) covariance between $\boldsymbol{f}_X$ and $f_y$.

From $p(\boldsymbol{f}_X,f_y)$ and using the multivariate normal theorem we may calculate the conditional probability distribution as
\begin{align}
    p(f_y|\boldsymbol{f}_x)=\mathcal{N}\left(\boldsymbol{\mu}_{\rm interp},K_{\rm interp}\right)\\
\intertext{with}
\boldsymbol{\mu}_{\rm interp}=K_{Xy} K_y^{-1}\boldsymbol{f}_X\\
K_{\rm interp}=K_X-K_{Xy} K_y^{-1}K_{Xy}^T
\end{align}

In these expressions, $\boldsymbol{f}_X$ is known and we need to choose a form for the covariances. For this analysis where we expect 
the interpolated surface to vary smoothly, we choose the covariance (also called kernel) $K$ to be a linear combination of a squared exponential (or radial basis function) and a linear kernel:
\begin{equation}
    K_{i,j}=\sigma_{\rm rbf} e^{-\frac{(\boldsymbol{x}_i-\boldsymbol{x}_j)^2}{l_{\rm rbf}}}+\delta_D(i-j)\sigma_{\rm fit},
\end{equation}
where $K_{i,j}$ is the covariance between point $\boldsymbol{x}_i$ and $\boldsymbol{x}_j$, $\sigma_{\rm rbf}$ represents the amplitude of the variance, $l_{\rm rbf}$ the characteristic interpolation length-scale 
and $\sigma_{\rm fit}$, the possible error in the determination of $\boldsymbol{f}_X$.

In the Gaussian Process package that we use~\citep{gpy}, these three hyper-parameters are common to the five cosmological parameters, therefore we scale parameters by the standard deviation of current points X. The hyper-parameters are let free and 
fitted by maximizing the marginal log-likelihood of the probability function of the known points \citep{gp}:
\begin{equation}
    \log p(\boldsymbol{f}_X|X,K)=-\frac{1}{2}\boldsymbol{f}_X^TK^{-1}\boldsymbol{f}_X-\frac{1}{2}\log|K|-\frac{N}{2}\log(2\pi),
\end{equation}
with $N$, the number of points of the training sample.

In a five-dimensional parameter space, even an algorithm as fast as the Gaussian process interpolation would take too long to run on a dense grid spanning the parameter space, therefore we run a Markov Chain Monte Carlo (MCMC) to obtain the interpolated log-likelihood hypersurface at each iteration step.

One may further include additional priors when performing such sampling of the Gaussian process interpolation. In this analysis (see section~\ref{sec:data_analysis}), we study the impact of Gaussian priors: a prior on $\omega_b=0.0222\pm 0.0005$ inspired by the Big Bang Nucleosynthesis (BBN) analysis, and a prior on $n_s=0.96\pm 0.02$, which 
are the minimal priors used in \cite{eboss}.

\subsection{New iteration and convergence}
\label{sec:iteration}
To check that the Gaussian process prediction has converged, we calculate new model power spectra. The choice of new points is done using an acquisition function $A(\boldsymbol{x})$ that, in the general case, accounts for the interpolated log-likelihood and the error on the interpolation, $\sigma$:
\begin{equation}
\label{eq:acq}
    A(\boldsymbol{x})=\log L_{\rm interp}(\boldsymbol{x})+\alpha \sigma_{\log L_{\rm interp}}(\boldsymbol{x}))
\end{equation}
where $\alpha$ is a free parameter that we set to 0 in our standard analysis. 
This gives an acquisition function that just accounts for the value of the interpolated likelihood to maximize the time spent to refine the region of space with high likelihood. Tests performed with a non-zero value of $\alpha$ did not show any improvement.

The likelihood is computed at 10 points sampled from $A(\boldsymbol{x})$, which we add to previously computed points for a new Gaussian process regression; such procedure is repeated until convergence. 
This is tested by computing the Kullback-Leibler divergence between the estimated likelihood distributions 
at iterations $i$ and $i-1$, namely:
\begin{equation}
\begin{split}
    &D_{\rm KL}(\log L_{\rm interp, i}||\log L_{\rm interp, i-1})=\\
    &\frac{1}{2}\left[\log\frac{|C_{i-1}|}{|C_i|}-d+tr(C_{i-1}^{-1}C_i)+(\mu_{i-1}-\mu_i)^T C_{i-1}^{-1}(\mu_{i-1}-\mu_i)\right]
\end{split}
\end{equation}
where $d$ is the dimension of the parameter space. Consistently with \cite{pellejero}, we 
require the condition $D_{\rm KL}<0.1$ to be fulfilled to stop the iterative procedure.

\section{Results}
\label{sec:results}

In this section, 
we first study the efficiency of our analysis on mocks.
The accuracy of the TNS model as well as the impact of observational systematics were evaluated in~\cite{beutler,gilmarin,neveux} and are not repeated here.
Then, we present the results 
on BOSS and eBOSS data.

\subsection{Results on QSO mocks}

Our pipeline was tested on the mocks of section~\ref{sec:mocks} to evaluate the performance of our full 5-D cosmological parameter space analysis.

\subsubsection{Impact of the fiducial cosmology}
\begin{figure*}
\centering
	\includegraphics[width=0.85\textwidth]{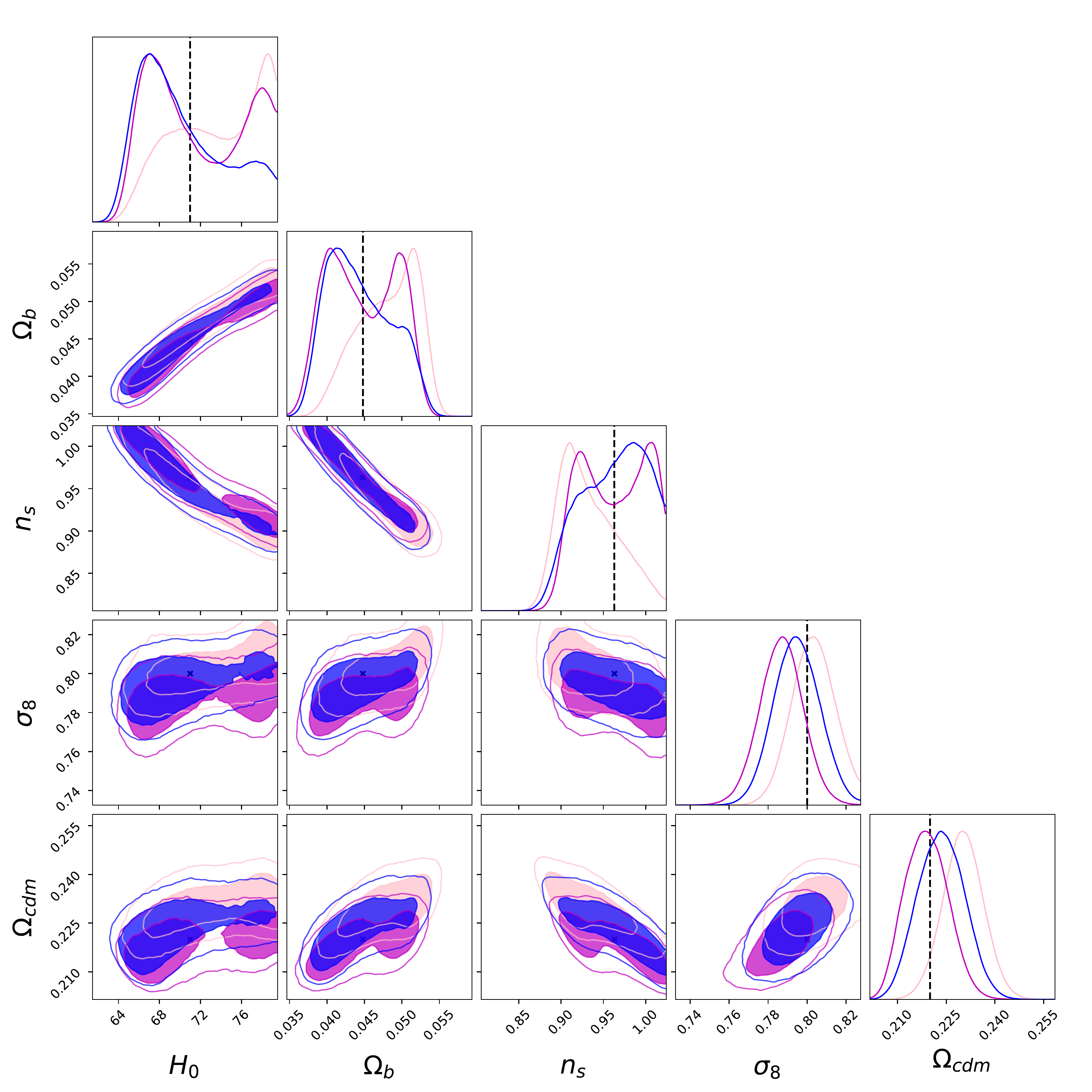}
    \caption{Constraints on cosmological parameters from fits to baseline QSO mocks. The blue contours represent the posteriors when the OuterRim cosmology is taken as the fiducial one, the purple contours (resp. pink) are the posteriors with the BOSS cosmology as the fiducial one, with (resp. without) updating the $k$-range.}
    \label{fig:mock_cosmo}
\end{figure*}

Although the power spectrum template is varied throughout the cosmological parameter space, we check the dependence of the cosmological constraints with the fiducial cosmology chosen to convert redshifts to distances.  Different tests were run to study this dependence, with results summarized in Table~\ref{tab:mock_gp_results} and figure~\ref{fig:mock_cosmo}.
We first take the OuterRim simulation cosmology (equation~\ref{eq:outerrim_cosmology}) as the fiducial cosmology. Note that the window function is also determined using that cosmology.
The Latin hypercube 
performs a sampling of the cosmological parameter space with 100 initial points. For each new iteration, 10 points are drawn using the acquisition function presented 
in section~\ref{sec:iteration}. The results of the process are presented in blue in figure~\ref{fig:mock_cosmo}. 
The posteriors encompass the parameter expected values, even for $H_0$, $\Omega_{\rm b}$ or $n_s$ which are not usually constrained by a clustering analysis. Moreover the $\Omega_{\rm cdm}$ and $\sigma_8$ parameters appear to be constrained with good accuracy and precision.

\begin{table*}
	\centering
	\caption{Constraints on the cosmological parameters in the QSO mock analysis for 
	different configurations. 
	The errors 
	are the 68\% marginalized confidence 
    intervals. For 
	$H_0$ 
	we report the 95$\%$ lower bound in the last three lines since the error interval reaches the flat prior boundaries we set in the fits. In the line "$\rm C_{fid} = C_{BOSS}~\&~k=[0.0209,0.3134]~\rm h\cdot Mpc^{-1}$", the results are presented as 
	95$\%$ confidence intervals for the $n_s$ and $\Omega_{\rm b}$ parameters because of 
	their bi-modal posteriors.}
	\label{tab:mock_gp_results}
	\hspace*{-1cm}
	\begin{tabular}{|l|c|c|c|c|c|} 
	\hline
    Configuration & $\Omega_{\rm cdm}$ & $\sigma_8$ & $n_s$ & $\Omega_{\rm b}$ & $H_0$ \\
\hline
\hline
OuterRim cosmology input values & 0.2200 & 0.800 & 0.963 & 0.0447 & 71.00\\
\hline
$\rm Cosmo_{ fid}=Cosmo_{OR}$ (baseline)& $0.2234\pm 0.0072$ & $0.794\pm 0.012$ & $0.986^{+0.016}_{-0.067}$ & $0.0412^{+0.0082}_{-0.0012}$ & $66.98^{+8.4}_{-0.87}$ \\
\hline
$\rm Cosmo_{ fid}=Cosmo_{BOSS}$ & $0.2297\pm 0.0061$ & $0.803\pm 0.011$ & $0.9095^{+0.061}_{-0.0099}$ & $0.05157^{+0.00060}_{-0.0079}$ & $> 66.48 (95\%)$ \\
\hline
$\rm C_{fid}=C_{BOSS}~\&~k=[0.0209,0.3134]~\rm h\cdot Mpc^{-1}$ & $0.2183\pm 0.0066$ & $0.787\pm 0.011$ & $0.898 < n_s^{\rm fit} < 1.020~(95\%)$ & $0.0374 < \Omega_{\rm b}^{\rm fit} < 0.0522~(95\%)$ & $> 65.40 (95\%)$ \\
\hline
alternative mock $\rm C_{fid}=C_{OR}$ & $0.226\pm 0.0067$ & $0.803^{+0.0089}_{-0.014}$ & $0.929^{+0.064}_{-0.011}$ & $0.0501^{+0.00064}_{-0.0081}$ & $> 66.09 (95\%)$ \\
\hline
\end{tabular}
\end{table*}

Secondly, we perform the same analysis using the BOSS cosmology as the fiducial one (equation~\ref{eq:fiducial_cosmology}). This choice is extreme as the BOSS cosmology ($\Omega_{\ rm m}^{\rm BOSS} = 0.31$) is 
significantly different (around 7 times the Planck error on that parameter~\citep{Planck18}) from that of the 
OuterRim simulation ($\Omega_{\rm m}^{\rm OR} = 0.2648$), which 
induces a dilation 
of $6.4\%$ and $3.5\%$, 
of the 
parallel and perpendicular to the line of sight distances, respectively. In the general case, 
power spectrum analyses are 
performed on a specified $k$ range. 
In different cosmologies, this implies that different physical modes enter the $k$ range under consideration. To overcome this issue, we modify the initial $k$ range (see Table~\ref{tab:k_par_survey}) by scaling the limits of the range by a factor that depends on the isotropic distance scales as: 
\begin{equation}
\label{eq:rescaling}
    k^{\rm BOSS}= k^{\rm OR}\frac{D_{\rm V}^{\rm OR}}{D_{\rm V}^{\rm BOSS}}
\end{equation}
where the distances are in $\rm [h^{-1}\cdot Mpc]$ to also account for the different values of $H_0$ between the two cosmologies. In the present case, the rescaled range is therefore $k \in [0.0209,0.3134]~\rm h\cdot Mpc^{-1}$.

The purple contours and posteriors in figure~\ref{fig:mock_cosmo} show the results using the BOSS cosmology and the rescaled $k$-range. For all parameters, the results 
agree reasonably well with those obtained with the true cosmology of the simulation.
The marginalized 
contour in the $\sigma_8-\Omega_{\rm cdm} $ plane shows a slight 
drift along the line of degeneracy, 
equivalent to $0.9\% $ and $2.3\% $ shifts 
for $\sigma_8$ and $\Omega_{\rm cdm}$, respectively. This is to be compared to $1.1\% $ and $2.8\% $ without updating the $k$-range.
For the other three parameters, $H_0$, $\Omega_{\rm b}$ and $n_s$, 
we note that the posteriors are bi-modal. This is likely to be due to the fact that our covariance matrix is approximate since it is based on 100 non-independent realisations. 
Regardless of this issue, the marginalized 2-dimensional contours of the two analyses nicely overlap for these three parameters and their 68 and 95$\%$ constraints are in agreement.

We also performed this mock analysis 
using the BOSS cosmology as the fiducial one 
but with the initial range, $k \in [0.02,0.3]~\rm h\cdot Mpc^{-1} $. The pink contours 
in figure~\ref{fig:mock_cosmo} show the corresponding constraints. 
For $\sigma_8$ and $\Omega_{\rm cdm}$, the difference in contours and posteriors due to different $k$ ranges is larger than that due to different fiducial cosmologies.
This tends to prove that taking into account the same physical modes is of prime importance to obtain similar constraints.
 For the other parameters, the updating $k$-range makes the contours and the likelihood profiles closer to those obtained with the true cosmology.
Finally, the choice of the fiduciary cosmology may amount to a choice of range in $k$, which should be marginalised in a real data analysis.

\subsubsection{Test using different HOD realizations}

\begin{figure*}
\centering
	\includegraphics[width=0.85\textwidth]{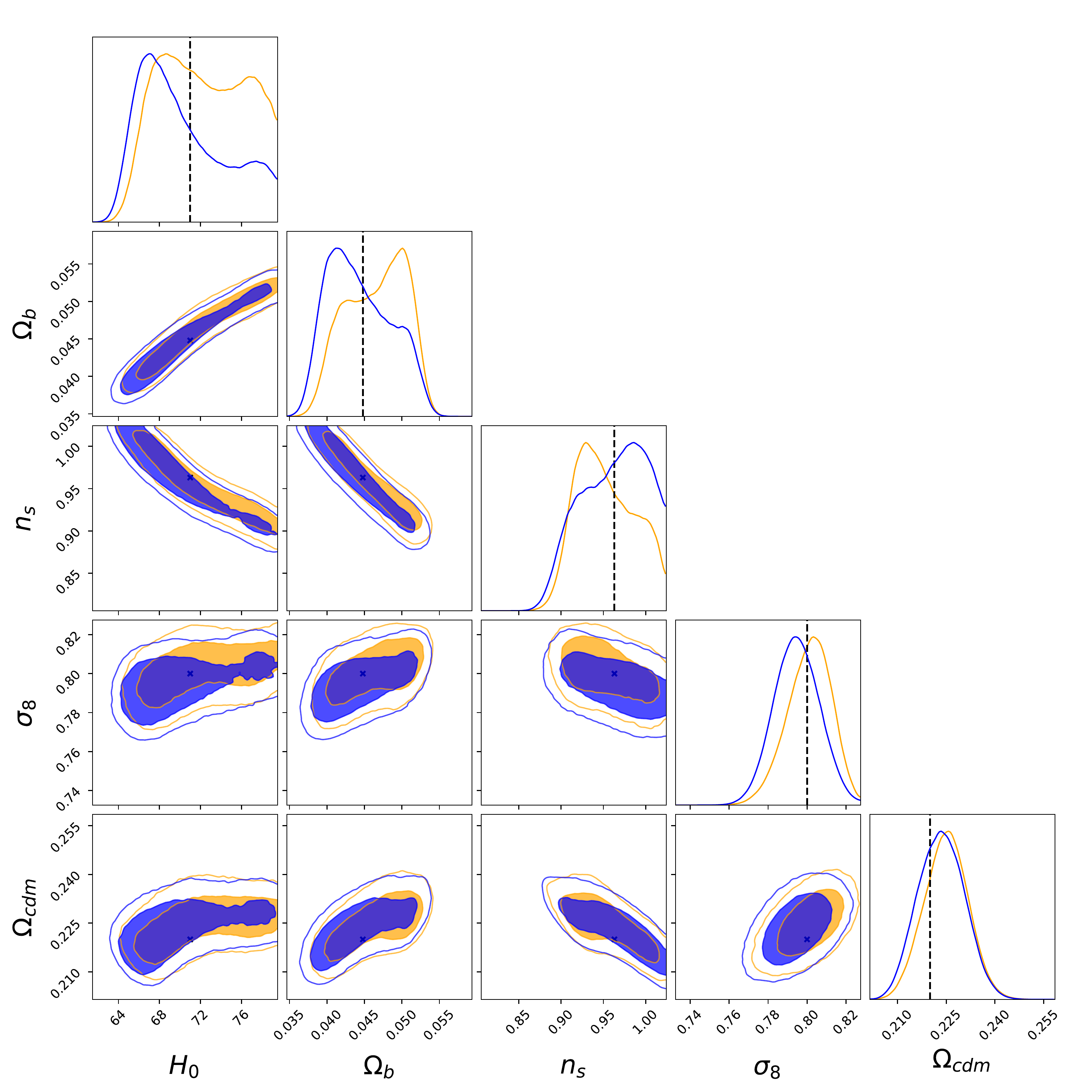}
    \caption{Constraints on cosmological parameters from fits to QSO mocks. The blue (resp. orange) contours represent the analysis posteriors for the baseline (resp. alternative) mocks. The difference being the HOD prescription. 
    All contours assume the OuterRim cosmology as the fiducial one.}
    \label{fig:mock_gp_hod}
\end{figure*}

In previous section, we use the mock2. We test, here, with another HOD prescription called mock4 analysed in the same way, 
with the OuterRim cosmology as the fiducial 
one.
Results are given in figure~\ref{fig:mock_gp_hod} and table~\ref{tab:mock_gp_results}. As for the baseline mocks, the posteriors for mock4 encompass the parameter values expected from the simulation cosmology.
As in the previous section, posteriors obtained from the two mocks are in good agreement, especially when considering that the covariance matrix used in the cosmological fit only encompasses noise in the galaxy - halo connection and not cosmic variance (see Section~\ref{sec:mocks}).
For 
$\sigma_8$ and $\Omega_{\rm cdm}$, we note a slight 
shift of the contour along the 
degeneracy line, which results in a difference of $1.1\%$ and $1.2\%$ on the two parameter best fit values, respectively.
 
 Summarising, these mock studies 
 demonstrate that the technique proposed here allows 
 to recover the baseline cosmological parameters at $1.5\% $ for $\sigma_8$ and $3\% $ for $\Omega_{\rm cdm}$ independently of the chosen fiducial cosmology.

\subsection{Results on data}
\label{sec:data_analysis}

The three data samples of section~\ref{sec:data} were analysed in the same way as mocks. 
Our main results are shown in figures~\ref{fig:gp_data_prior} and~\ref{fig:gp_data_surveys}  and summarised in table~\ref{tab:data_gp_results}.
We present hereafter combined results as well as results from the individual samples and discuss the impact of priors from external probes. 

\subsubsection{Combined survey analysis}


\begin{figure*}
\centering
	\includegraphics[width=0.9\textwidth]{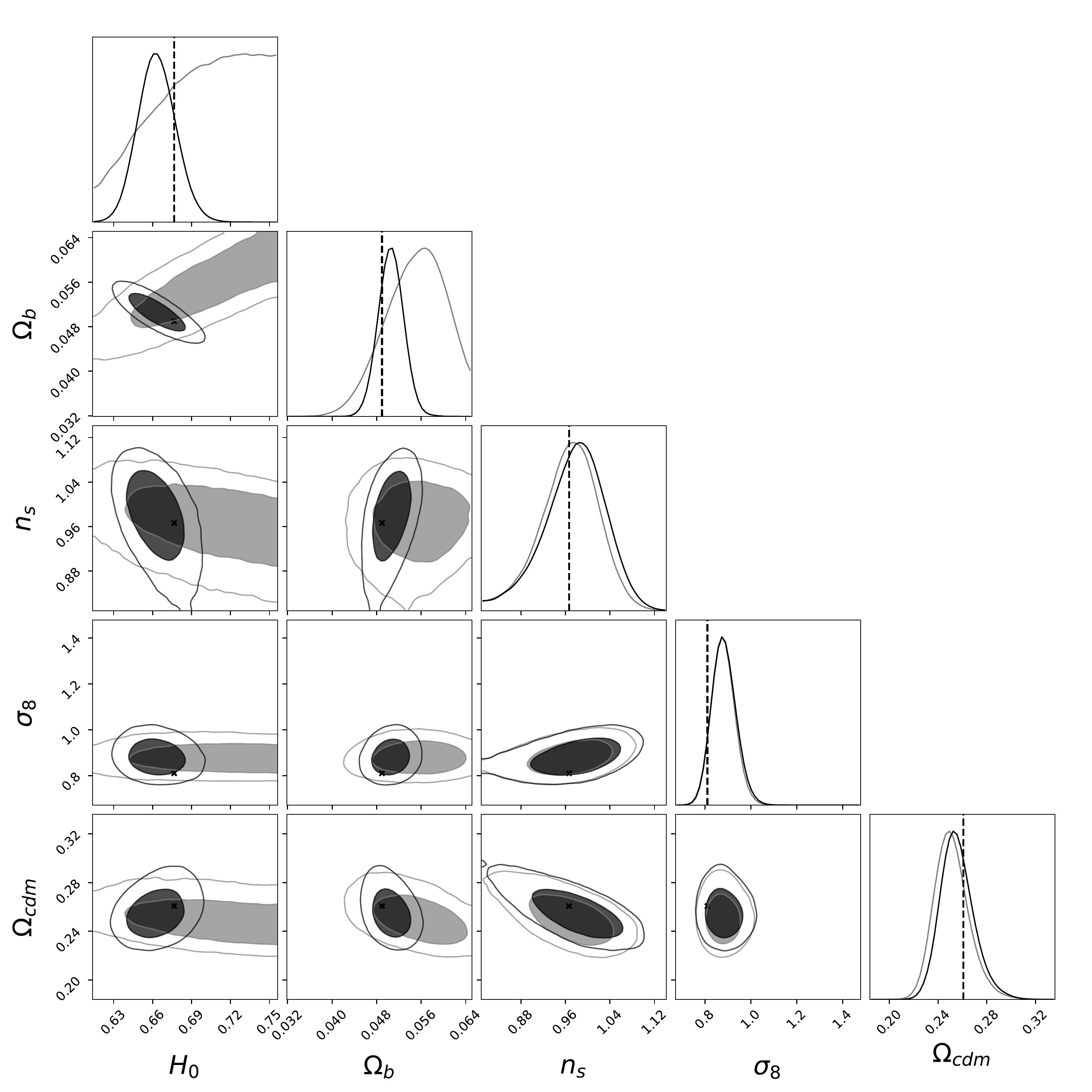}
    \caption{Constraints on the cosmological parameters for the three (e)BOSS surveys used in this analysis. The gray contours represent the posteriors without any external prior, while the black contours
    assume a prior on $\omega_b$ from BBN. The dashed line represents the $\Lambda$CDM Planck best fit.
    }
    \label{fig:gp_data_prior}
\end{figure*}

\begin{table*}
	\centering
	\caption{Constraints 
	on cosmological parameters 
	from the analysis of BOSS low-z galaxies, 
	and eBOSS LRG and QSO samples. The errors 
	are the 68\% marginalized confidence 
	intervals. 
	 $\Omega_{\rm m}$ is a derived parameter.}
	\label{tab:data_gp_results}
	\begin{tabular}{|l|c|c|c|c|c|c|c|}
	\hline
    Data & Priors & $\Omega_{\rm cdm}$ & $\sigma_8$ & $n_s$ & $\Omega_{\rm b}$ & $H_0$ & $\Omega_{\rm m}$ \\
\hline
\hline
3 surveys & & $0.249^{+0.017}_{-0.011}$ & $0.877 \pm 0.049$ & $0.973\pm 0.054$ & $0.0566^{+0.0040}_{-0.0074}$ & & $0.304^{+0.016}_{-0.010}$\\
\hline
3 surveys & $\omega_{\rm b}$ & $0.253^{+0.017}_{-0.010}$ & $0.877\pm 0.051$ & $0.983\pm 0.056$ & $0.0506\pm 0.0022$ & $66.1\pm 1.5$ & $0.304^{+0.016}_{-0.010}$ \\
\hline
3 surveys & $\omega_{\rm b}~\&~n_s$ & $0.256\pm 0.010$ & $0.869\pm 0.046$ & $0.963\pm 0.020$ & $0.0503\pm 0.0020$ & $66.4\pm 1.4$ & $0.308\pm 0.010$ \\
\hline
LRG eBOSS + galaxies low-z BOSS & & $0.285\pm 0.021$ & $0.721^{+0.060}_{-0.028}$ & $< 0.978 (95\%)$ & $0.0423^{+0.0090}_{-0.0047}$ & & $0.328\pm 0.018$ \\
\hline
LRG eBOSS + galaxies low-z BOSS & $\omega_{\rm b}$ & $0.277\pm 0.018$ & $0.738^{+0.056}_{-0.031}$ & $< 0.988 (95\%)$ & $0.0483\pm 0.0025$ & $67.9\pm 1.8$ & $0.324\pm 0.017$ \\
\hline
LRG eBOSS + galaxies low-z BOSS & $\omega_{\rm b}~\&~n_s$ & $0.265\pm 0.013$ & $0.763\pm 0.046$ & $0.953\pm 0.019$ & $0.0490\pm 0.0024$ & $66.9\pm 1.9$ & $0.315\pm 0.013$ \\
\hline
QSO eBOSS & & $0.243^{+0.027}_{-0.014}$ & $1.15\pm 0.10$ & $1.043^{+0.033}_{-0.075}$ & $> 0.0472 (95\%)$ & & $0.304^{+0.025}_{-0.015}$ \\
\hline
QSO eBOSS & $\omega_{\rm b}$ & $0.252^{+0.032}_{-0.014}$ & $1.13\pm 0.11$ & $1.003^{+0.046}_{-0.100}$ & $0.0527\pm 0.0033$ & $64.6\pm 2.0$ & $0.306^{+0.030}_{-0.015}$ \\
\hline
QSO eBOSS & $\omega_{\rm b}~\&~n_s$ & $0.267\pm 0.016$ & $1.12\pm 0.10$ & $0.967\pm 0.021$ & $0.0520\pm 0.0031$ & $65.1\pm 1.9$ & $0.321\pm 0.016$ \\
\hline
\end{tabular}
\end{table*}

As shown in figure~\ref{fig:gp_data_prior}, when no external priors are used (grey contours), we find that $H_0$ is very little constrained, as expected in a galaxy clustering analysis. 
Constraints can be set on $\Omega_{\rm b}$ and $n_s$, namely $\Omega_{\rm b} = 0.0566^{+ 0.0040}_{- 0.0074}$ and $n_s = 0.973\pm 0.054$. The uncertainty is an order of magnitude larger than that of the Planck CMB analysis~\citep{Planck18}, 
$\Omega_{\rm b}^{\rm Planck} = 0.0493\pm 0.0003$ and $n_s^{\rm Planck} = 0.965\pm 0.004$. However, galaxy clustering analyses do not usually report any constraint on these parameters.
The value of $\Omega_{\rm cdm}=0.249^{+0.017}_{-0.011}$ 
is within $1\sigma$ of the Planck value. 
This is true also for the other parameters, except $\sigma_8=0.877\pm 0.049$ 
whose value is $1.4\sigma$ above that of Planck.

 Including a prior from the BBN (see section~\ref{sec:gp}) on the value of $\omega_{\rm b}=\Omega_{\rm b}\cdot \rm h^{2}$ (black contours), 
the constraints improve on $H_0=66.1\pm 1.5~\rm km.s^{-1}.Mpc^{-1}$ and $\Omega_{\rm b}=0.0506\pm 0.0022$. It is however interesting to note that this prior has very little impact on the other constraints 
e.g.
$\Omega_{\rm cdm}=0.253^{+0.017}_{-0.010}$ and $n_s=0.983\pm 0.056$ 
or no impact at all, e.g. $\sigma_8=0.877\pm 0.051$ and $\Omega_{\rm m}=0.304^{+0.016}_{-0.010} $.
The matter density 
is identical, only its share between baryons and cold dark matter has varied.

Adding a prior on $n_s$ (see section~\ref{sec:gp}) to the previous analysis (black contours in figure~\ref{fig:gp_data_surveys}) 
induces a slight change which remains very small with respect to the statistical error. Altogether, the weak impact of the priors shows that 
robust cosmological constraints on $\Omega_{\rm m}$ and $\sigma_8$ can be obtained without any additional information from external probes on parameters not constrained by galaxy clustering analyses like $n_s$ and $\omega_{\rm b}$.

\begin{figure*}
\centering
   \includegraphics[width=0.49\textwidth]{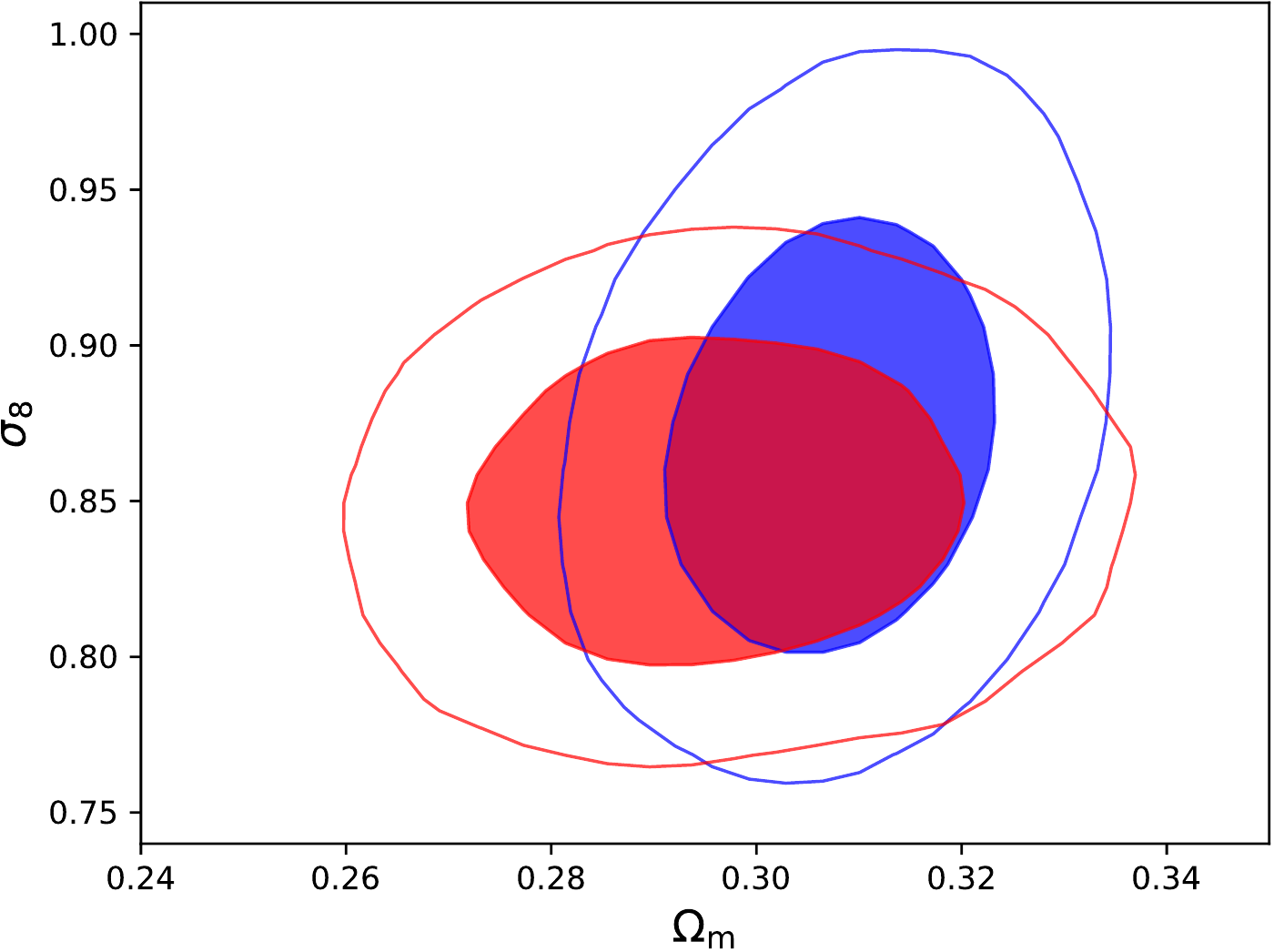}
   \includegraphics[width=0.49\textwidth]{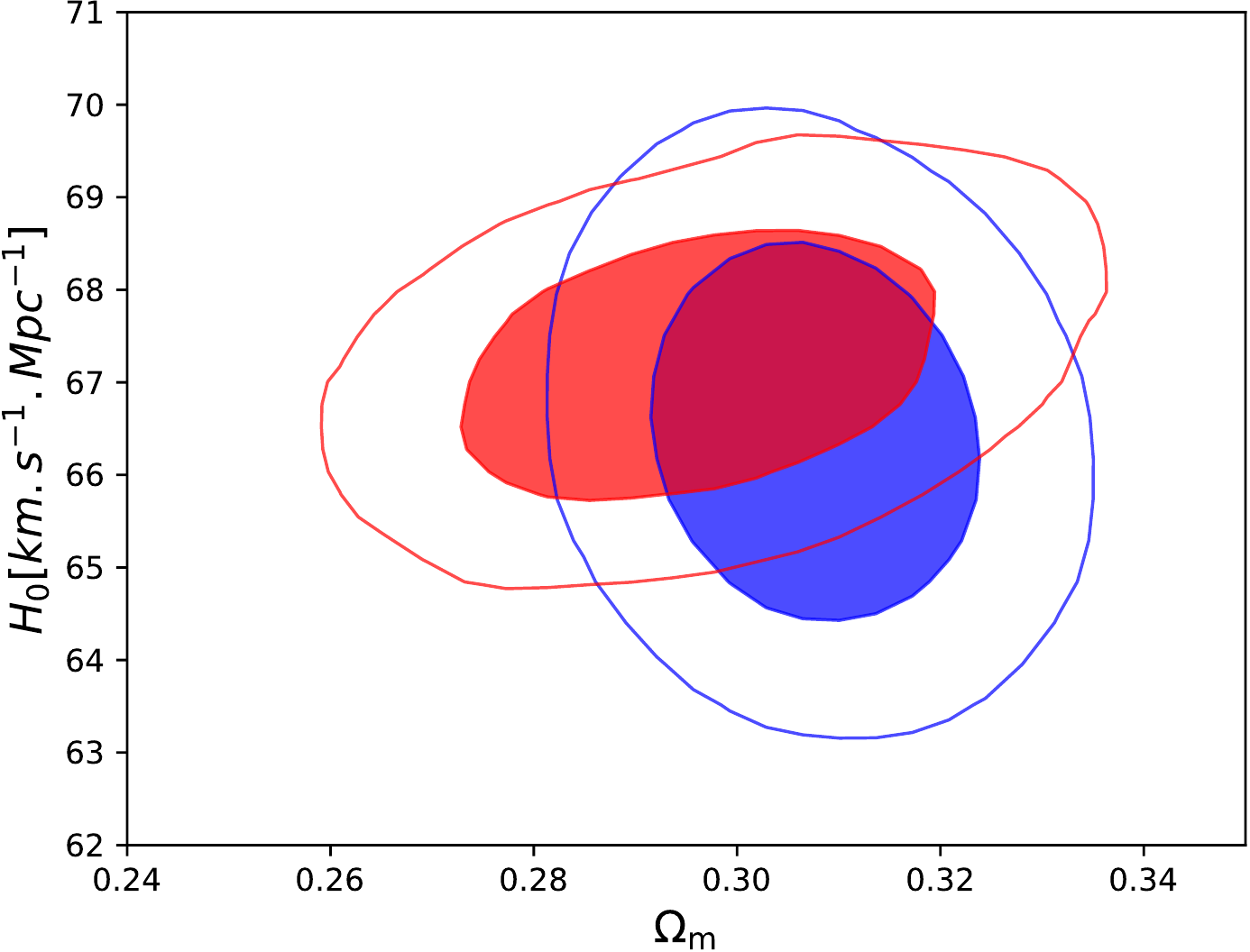}
   \caption{Constraints on the cosmological parameters from this study in blue and from the standard cosmological analysis of \citep{eboss} in red. In both cases, the same priors are used on $\omega_{\rm b}$ and $n_s$. However, it is important to note that we use the eBOSS QSO and LRG surveys as well as the BOSS low-z galaxy survey, while the standard analysis uses all SDSS surveys, which adds the MGS sample, the BOSS high-z galaxies and the eBOSS ELG and Lyman-$\alpha$ surveys. The left-hand plot shows the $\sigma_8-\Omega_{\rm m}$ plane, that on the right the $H_0-\Omega_{\rm m}$ plane.
   }
    \label{fig:sdss_vs_gp}
\end{figure*}

In figure~\ref{fig:sdss_vs_gp}, we compare our results with 
those presented in \cite{eboss} 
based on standard clustering analysis techniques 
and the $\Lambda$CDM framework. This comparison is done under the same assumptions, i.e. including the same priors on $\omega_{\rm b}$ and $n_s$. 
The contours have similar areas but different orientations and the constraints projected onto $\sigma_8$ and $H_0$ are stronger with the standard analysis, most probably because of the more extended range in redshift of the extra samples that it includes, as detailed below.
The present analysis allows a better constraint on $\Omega_{\rm m}$ despite the differences in terms of 
number of objects and effective volume. This stronger constraint illustrates the significant gain in precision 
allowed by a varying template 
analysis with no compression of the information. To emphasize this, we recall that the present analysis is composed of three samples (BOSS low-z galaxies, 
eBOSS LRGs and quasars) 
while the standard analysis is composed, in addition, of the MGS sample, BOSS high-z galaxies, 
and eBOSS ELG and Lyman-$\alpha$ samples. 
Therefore, the complete SDSS survey contains 2.5 million objects over a redshift range of $0.07<z<4$ while this analysis includes 1.3 million objects 
over $0.2<z<2.2$. In addition, the standard analysis uses a consensus between 
the correlation function and 
power spectrum analyses, which usually improves errors by $~10\%$, while the present study is only carried out in Fourier space.

On the other hand, in this analysis, we do not include errors related to systematic effects. 
Nevertheless, systematic observational errors have been investigated in individual power spectrum analyses for the informative parameters ($\alpha_{\perp,\parallel}$ and $f\sigma_8$) and remain small compared to the statistical error 
(e.g. $\sigma^{\rm obs}/\sigma^{\rm stat}\sim 0.1$ for quasars). 
Their addition in quadrature is therefore negligible. Considering the error due to the power spectrum modelling, we remind that the dominant source of error in eBOSS analyses comes from the specification of the cosmology for the model computation.
Taking the difference between our mock results obtained with the two different fiducial cosmologies (due to different effective $k$-ranges) as model systematic error leads to $\sigma^{\rm mod}_{\sigma_8}=0.012$ and $\sigma^{\rm mod}_{\Omega_{\rm cdm}}=0.007$,
which corresponds to $25\%$ and $44\%$ of the statistical error, respectively. 
 
\begin{figure*}
\centering
	\includegraphics[width=0.6\textwidth]{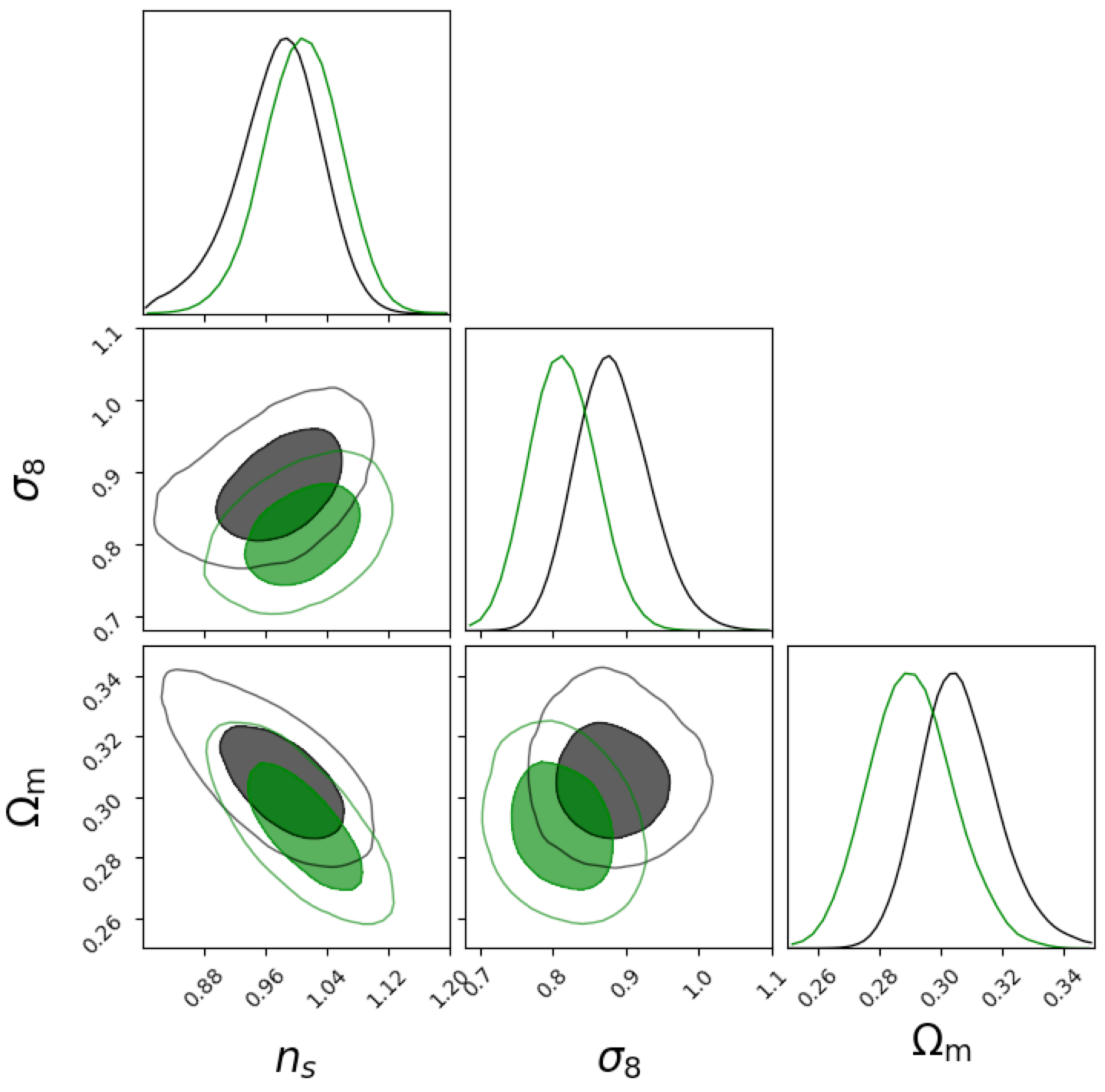}
    \caption{Cosmological constraints on $Omega_{\rm m}$, $\sigma_8$ and $n_s$. Comparison of our analysis with the BBN-inspired $\omega_{\rm b}$ prior (black contours) to the constraints obtained with an other full shape eBOSS analysis \citep{semenaite} (green contours). 
    }
    \label{fig:comparison_agne}
\end{figure*}

In addition to the present analysis, the full shape two point 
correlation function of the eBOSS QSO and BOSS DR12 galaxy samples sample were analysed by \cite{semenaite}. In their work, they 
directly constrain the cosmological parameters in configuration space by 
fitting their modelled two-point correlation function to the data and 
focus on $h$-independent parameters $\omega_{\rm m}$, $\omega_{\rm DE}$ 
and $\sigma_{12}$ as proposed by \cite{sanchez_s12}. In 
figure~\ref{fig:comparison_agne}, we compare their results with the ones 
obtained in this work. The three parameters that are recovered from both 
analyses ($\sigma_8$, $\Omega_{\rm m}$ and $n_s$) are found to be 
compatible at the 1 $\sigma$ level, although the samples considered in 
both analyses are not exactly the same (our work includes the eBOSS LRG 
sample which is not used in \cite{semenaite} who, however, additionally 
analyse the measurements from BOSS DR12 high-z sample). In addition to 
this, there are also differences in modelling the non-linear tracer 
power spectrum, in particular, \cite{semenaite} use a different model 
for non-linear clustering predictions as well as a different bias 
prescription. Finally, the two analyses differ in the choice of priors - 
firstly, in terms of parameters on which a flat prior is imposed and 
secondly, in the use of the BBN prior with this analysis employing (when 
applicable) a significantly tighter Gaussian prior instead of the wider 
flat informative prior used in \cite{semenaite}.

\subsubsection{Single tracer analysis}

\begin{figure*}
\centering
	\includegraphics[width=0.9\textwidth]{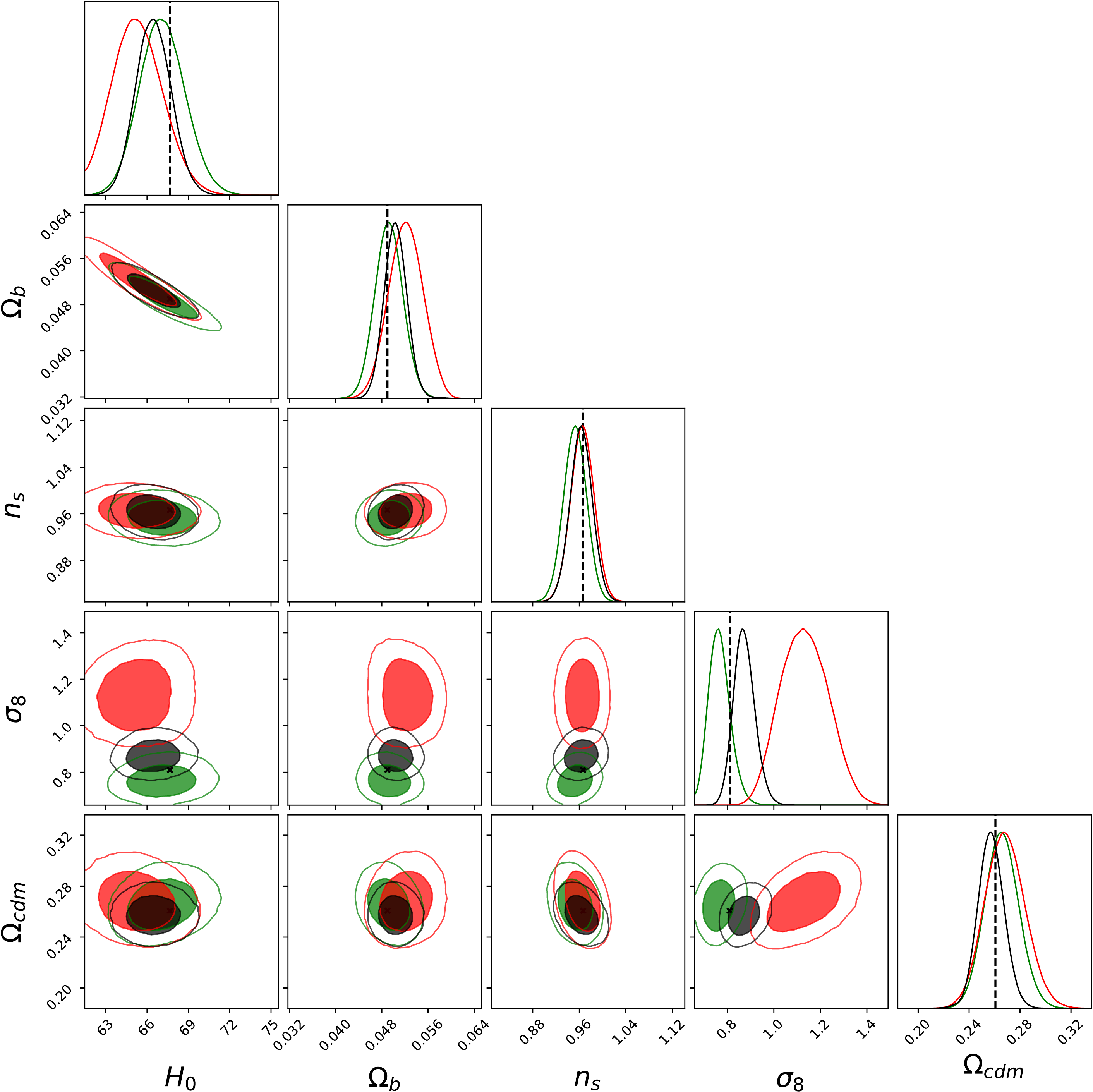}
    \caption{Constraints on the cosmological parameters for the three BOSS surveys by separating the constraints of quasars (in red), and galaxies (in green). The black contours represent the combined posterior of the three surveys. A prior on $ \omega_b$ and $n_s$ have been taken into account. The dashed line represents Planck's best fit. 
    }
    \label{fig:gp_data_surveys}
\end{figure*}

In this section, we analyse the low redshift samples (BOSS galaxies and eBOSS LRGs) and the high redshift sample (eBOSS quasars) separately. Figure~\ref{fig:gp_data_surveys} presents the different contours when including the same priors on $\omega_{\rm b} $ and $n_s$ as for the combined analysis. 
Indeed, as written in table~\ref{tab:data_gp_results},
without these priors, $n_s$ (resp. $\omega_b$) is not constrained better than its range of variation used in the fit to the galaxy (resp. quasar) sample. This may be due to the lower sample size and to the loss of leverage between high and low redshifts in individual analyses compared to the combined one.
Including priors in the individual analyses has thus a more important impact than on the combined one. However, this difference is still smaller than the statistical uncertainty. 

The posteriors of the low and high redshift analyses agree on $\Omega_{\rm cdm}$ and show a small but not negligible difference on
$H_0$, $\Omega_{\rm b}$ and $n_s$, 
corresponding to deviations of $0.9\sigma$, $1\sigma$ and $0.7\sigma$, respectively. 
The constraints on these parameters are also in agreement with Plank results, and so do our derived $\Omega_{\rm m}$ results. 
On the other hand, the galaxy and quasar analyses differ by $3.6\sigma$ on $\sigma_8$. The constraint from the galaxy sample agrees with Planck result ($1.6\sigma$ lower) while that from the quasar sample is $3.1\sigma$ higher. A $2\sigma$ bias with respect to Planck is also visible in the standard analysis \citep{neveux}, when fixing the linear growth rate of structure to the GR expectation. Nevertheless, this discrepancy is reinforced in the present study.

This notably high $\sigma_8$ value from high redshift quasars constitutes an unexplained tension with the other data sets used here. 
This may be due to an unknown or poorly understood systematic effect, physics not explained by the 
$\Lambda $CDM model, or simply a statistical fluctuation. 
We did a few tests to check the robustness of this difference. In our framework, we constrain $\sigma_8(z=0)$ from the measurement of $f\sigma_8(z)$ at $f$ fixed to the value predicted by general relativity at each point of the cosmological parameter space. However, the $\sigma_8$ parameter also appears in the amplitude of the power spectrum monopole through the product $b_1\sigma_8(z)$ (in linear theory). 

\begin{figure*}
\centering
	\includegraphics[width=0.7\textwidth]{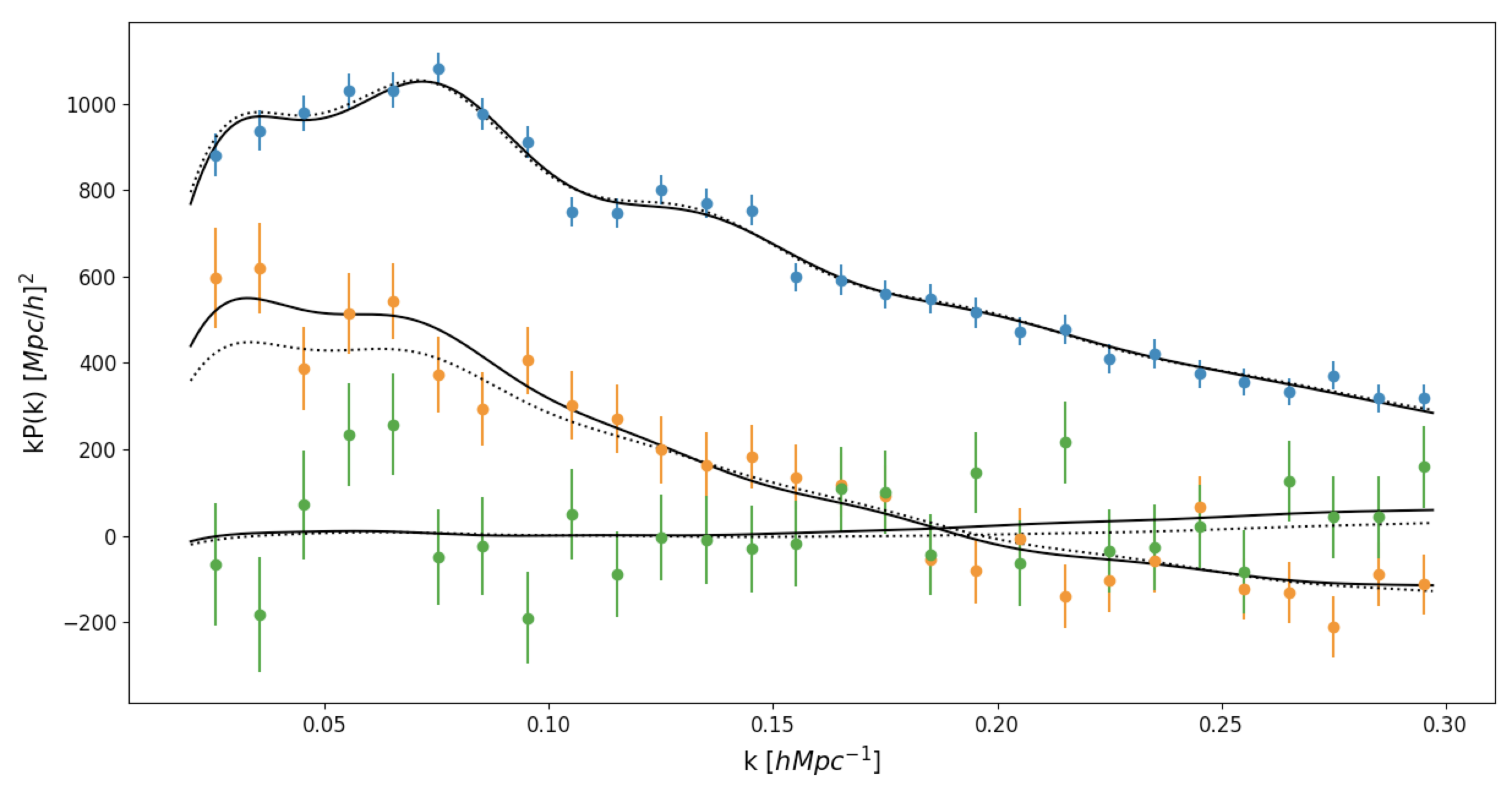}
    \caption{Comparison of the eBOSS NGC data quasar power spectrum multipoles with 
    best fit predictions from 
    the eBOSS quasar sample analysis (solid line) and from the combined analysis (dotted line), both with no external priors.
    The corresponding best fit cosmologies 
    are $h = 0.7, \,\, \Omega_{\rm cdm} = 0.252, \,\, \Omega_{b} = 0.0640, \,\, \sigma_{8} = 1.150, \,\, n_s=1.043$ for the solid line and 
    $h = 0.7, \,\, \Omega_{\rm cdm} = 0.249, \,\, \Omega_{b} = 0.0566, \,\, \sigma_{8} = 0.877, \,\, n_s=0.973$ for the dotted line.}
    \label{fig:ps_fit}
\end{figure*}

In figure~\ref{fig:ps_fit}, we show the power spectrum multipoles for the NGC part of the quasar sample (the SGC shows a similar behavior) along with the best fit model for the cases where all tracers (dotted line) or only the quasar sample (solid line) are considered (no external priors). For both fits, there is no constraining power on the value of $h$ and we have set it to $h=0.7$ for the sake of comparison. The monopole is unaffected as a consequence of the $b_1\sigma_8$ degeneracy previously mentioned. On the other hand, the amplitude of the best fit model for the quadrupole at large scales depends upon the samples considered. Therefore, the difference in $\sigma_8$ between the galaxy and quasar fits stems mostly from the amplitude of the quadrupole measured with the eBOSS quasar sample. More data from upcoming surveys should help settle this issue.
We verified that the updated scheme to mitigate photometric systematics using the neural network approach proposed by \cite{rezaie} and used in the $f_\mathrm{NL}$ analysis of~\cite{mueller} had no effect on the power spectrum multipoles in the $k$-range of interest of this work and cannot account for the high value of $\sigma_8$.
The study presented in \cite{semenaite} also considers the quasar sample alone and their results differ from ours by 1.6 and 1$\sigma$ for $\sigma_8$ and $\Omega_{\rm m}$ respectively, which differences in the analysis may explain.


\section{Conclusions}

We analysed the power spectrum of a subsample of the BOSS and eBOSS 
data using a TNS model with 2-loop RegPT correction terms.
To avoid the fiducial cosmology dependency of the clustering analysis, we 
built an iterative emulator of the analysis likelihood surface by varying the cosmology of the template model. We fit 
model power spectra for different cosmologies to the data power spectrum, 
fixing the dilation scales 
and the growth rate of structure to their values expected in each tested cosmology
and 
letting the nuisance parameters vary freely. Then, we reconstruct the full likelihood surface in the 5D cosmological parameter space using a Gaussian process algorithm. This technique allows us to fit the cosmological parameters without the standard compression in $f\sigma_8$ and scaling parameters $\alpha_\parallel$ and $\alpha\perp$ and with minimal power spectrum model computation.

We tested this pipeline on mocks built from the OuterRim N-body simulation and efficiently recovered the five cosmological parameters with good accuracy and precision. Then, we analysed three samples of the SDSS spectroscopic surveys both individually and jointly, namely the BOSS low-z galaxy sample 
and the eBOSS LRG and QSO samples. 
The analysis of the QSO sample 
leads to a $\sigma_8$ value significantly different from 
that predicted from the Planck constraints~\citep{Planck18}. This bias is also visible in the standard analysis \citep{neveux}, with a $2\sigma$ deviation from the Planck analysis when fixing the linear growth rate of structure to the GR expectation. Nevertheless, this discrepancy is reinforced in the present study and reaches a $3.1\sigma$ significance. All other parameters are in good agreement with the constraints from the CMB analysis.

The combined likelihood of the three samples 
allows us to fit $\Omega_{\rm b}$, $n_s$, $\Omega_{\rm m}$ and $\sigma_8$ without any external prior. We compare our final results with the cosmological analysis of the full SDSS survey; we obtain similar constraints on the $H_0$, $\sigma_8$ and $\Omega_{\rm m}$ parameters using 
data sample twice as small. The $\Omega_{\rm m}$ parameter is even better constrained due to the information loss in the compression step in the standard analysis.
Such a pipeline that removes the dependence in the fiducial cosmology at all stages of the analysis could be used in future spectroscopic surveys such as DESI or Euclid to reduce the systematic and statistical errors on the cosmological parameters.

\section*{Acknowledgements}

R. Neveux acknowledges support from grant ANR-16-CE31-0021, eBOSS and from ANR-17-CE31-0024-01, NILAC. 

Funding for the Sloan Digital Sky Survey IV has been provided by the Alfred P. Sloan Foundation, the U.S. Department of Energy Office of Science, and the Participating Institutions. SDSS acknowledges support and resources from the Center for High-Performance Computing at the University of Utah. The SDSS web site is www.sdss.org.

SDSS is managed by the Astrophysical Research Consortium for the Participating Institutions of the SDSS Collaboration including the Brazilian Participation Group, the Carnegie Institution for Science, Carnegie Mellon University, Center for Astrophysics | Harvard \& Smithsonian (CfA), the Chilean Participation Group, the French Participation Group, Instituto de Astrofísica de Canarias, The Johns Hopkins University, Kavli Institute for the Physics and Mathematics of the Universe (IPMU) / University of Tokyo, the Korean Participation Group, Lawrence Berkeley National Laboratory, Leibniz Institut für Astrophysik Potsdam (AIP), Max-Planck-Institut für Astronomie (MPIA Heidelberg), Max-Planck-Institut für Astrophysik (MPA Garching), Max-Planck-Institut für Extraterrestrische Physik (MPE), National Astronomical Observatories of China, New Mexico State University, New York University, University of Notre Dame, Observatório Nacional / MCTI, The Ohio State University, Pennsylvania State University, Shanghai Astronomical Observatory, United Kingdom Participation Group, Universidad Nacional Autónoma de México, University of Arizona, University of Colorado Boulder, University of Oxford, University of Portsmouth, University of Utah, University of Virginia, University of Washington, University of Wisconsin, Vanderbilt University, and Yale University.

\section*{Data Availability}
The power spectrum, covariance matrices, and resulting likelihoods for cosmological parameters are available via the SDSS Science Archive Server (https://sas.sdss.org/)


\bibliographystyle{mnras}
\bibliography{references}



\appendix


\bsp	
\label{lastpage}
\end{document}